\documentclass[12pt]{article}

\usepackage{newtxtext,newtxmath}

\usepackage{graphicx}
\usepackage{lineno}

\usepackage[letterpaper,margin=1in]{geometry}

\renewenvironment{abstract}
	{\quotation}
	{\endquotation}

\date{}

\usepackage[labelsep=period]{caption}
\makeatletter
\renewcommand{\fnum@figure}{\textbf{Fig. \thefigure}}
\renewcommand{\fnum@table}{\textbf{Table \thetable}}
\makeatother

\usepackage{scicite}

\usepackage{url}
\usepackage{soul}

\def\scititle{
	Coherent terahertz control of metastable magnetization in FePS$_3$
}
\title{\bfseries \boldmath \scititle}

\author{
	Batyr Ilyas$^{1,\dagger}$,
	Tianchuang Luo$^{1,\dagger}$,
    Honglie Ning$^{1,\dagger}$,
    Emil Viñas Boström$^{2, 3}$, \and
    Alexander von Hoegen$^{1}$,
    Jaena Park$^{4}$, 
    Junghyun Kim$^{4}$,
    Je-Geun Park$^{4}$, \and
    Angel Rubio$^{2, 3, 5}$,
	Nuh Gedik$^{1, \ast}$\and
	\small$^{1}$Department of Physics, Massachusetts Institute of Technology, Cambridge, 02139, USA.\\
	\small$^{2}$Max Planck Institute for the Structure and Dynamics of Matter, Hamburg, 22761, Germany. \\
    \small$^{3}$Nano-Bio Spectroscopy Group, Departamento de Fisica de Materiales, Universidad del Pais Vasco, \\ \small San Sebastian, 20018, Spain. \\
    \small$^{4}$Department of Physics and Astronomy and Institute of Applied Physics, Seoul National University,\\ \small Seoul, 08826, Republic of Korea.\\
    \small$^{5}$ Initiative for Computational Catalysis, The Flatiron Institute, New York, NY, 10010, USA. \and
	\small$^\ast$Corresponding author. Email: gedik@mit.edu\and 
	\small$^\dagger$These authors contributed equally to this work.
}

\begin{document} 

\maketitle
\begin{abstract} \bfseries \boldmath
The crystal lattice governs the emergent electronic, magnetic, and optical properties of quantum materials, making structural tuning through strain, pressure, or chemical substitution a key approach for discovering and controlling novel quantum phases. Beyond static modifications, driving specific lattice modes with ultrafast stimuli offers a dynamic route for tailoring material properties out of equilibrium. However, achieving dynamic coherent control of the nonequilibrium phases via resonant excitation of lattice coherences remains largely unexplored. Such manipulation enables non-volatile, on-demand amplification and suppression of order parameters on femtosecond timescales, necessary for next-generation optoelectronic ultrafast computation.
In this study, we demonstrate coherent phononic control of a newly discovered, light-induced metastable magnetization in the van der Waals antiferromagnet FePS$_3$. By using a sequence of terahertz (THz) pulses, we modulate the magnetization amplitude at the frequencies of phonon coherences, whose infrared-active nature and symmetries are further revealed by polarization- and field-strength-dependent measurements. Furthermore, our two-dimensional THz spectroscopy, in tandem with first-principles numerical simulations, shows that these phonons nonlinearly displace a Raman-active phonon, which induces the metastable net magnetization. 
These findings not only clarify the microscopic mechanism underlying the metastable state in FePS$_3$ but also establish vibrational coherences in solids as a powerful tool for ultrafast quantum phase control, enabling manipulation of material functionalities far from equilibrium.
\end{abstract}


\noindent
The crystal structure is a key factor in determining the emergent phenomena in quantum materials. Control of lattice arrangements has proven to be a powerful route to discovering new phases. For instance, in two-dimensional layered materials, the introduction of long-wavelength superlattice periodicity through twisting or lattice mismatch has unveiled a rich variety of correlated electronic orders, including superconductivity \cite{cao2018unconventional, chen2019signatures} and fractional quantum anomalous Hall states \cite{cai2023signatures, zeng2023thermodynamic, park2023observation}. In magnetic systems, changes in the stacking configuration of atomic layers can similarly lead to diverse magnetic ground states \cite{song2021direct, xie2022twist}. More broadly, lattice tuning through established methods such as external pressure \cite{torikachvili2008pressure,mathur1998magnetically} and epitaxial strain \cite{boschker2017quantum, McLeod2020Multi-messengerManganite} has long been effective in stabilizing phases inaccessible under ambient conditions. Beyond these static approaches, dynamic structural control through selective excitation of phonon resonances has recently emerged as a promising approach to inducing novel nonequilibrium phases \cite{rini2007control, disa2021engineering}. This technique has enabled the realization of symmetry-broken phases that either have no equilibrium analogs \cite{nova2019metastable, li2019terahertz, ilyas2024terahertz} or are stabilized at temperatures exceeding their thermodynamic limits \cite{disa2023photo, rowe2023resonant}. Specifically, the concept of dynamical control of magnetism via coherent lattice excitations has been established by recent pioneering works \cite{nova2017effective, juraschek2017dynamical, radaelli2018breaking, afanasiev2021ultrafast, stupakiewicz2021ultrafast, Disa2020PolarizingField, disa2023photo}. This rapidly evolving direction of phase manipulation holds great promise for ultrafast and energy-efficient functional applications.

%

Despite these advances, many aspects of phononic control protocols remain poorly understood. One significant yet underutilized advantage is that phonons exhibit exceptionally longer coherence times compared to electrons~\cite{maccabe2020nano, zhang2023assessing}, which opens up numerous possibilities for inducing, sustaining, and modifying out-of-equilibrium states. Previous studies have demonstrated coherent modulation of phase transitions through phonon modes launched by near-infrared (NIR) pulses \cite{horstmann2020coherent, maklar2023coherent}. However, these approaches generate substantial light-induced heating due to the excessive energy deposited into the electronic subsystem and indiscriminately excite multiple Raman-active modes of different symmetries \cite{yan1985impulsive}. Such heating and lack of selectivity inherently hinder the on-demand phononic manipulation of fragile, temperature-sensitive phases. To date, coherent control of phase transitions through selective and resonant excitation of phonon modes with minimized heating has not been realized. 

To place this challenge in a broader context, while coherent control is well-established in molecular chemistry, its application to switch entire macroscopic quantum phases in solid-state systems is fundamentally distinct and challenging. Unlike isolated molecules, solids possess a continuum of electronic and phononic excitations that act as a thermal bath, typically inducing decoherence.  However, if such an environment can be properly navigated, crystalline solids may provide unique advantages. For example, using resonant terahertz excitation to isolate a specific, phase-preserving lattice pathway, can allow coherent control durations far exceeding the transient femtosecond timescales of molecules. Therefore, demonstrating such a coherence can survive long enough to deterministically dictate a macroscopic order parameter represents a fundamental leap from local molecular manipulation to the long-lived dynamic control of quantum materials.

Achieving this functionality requires meeting two conditions. First, an infrared (IR)-active phonon, resonantly excited with minimal disturbance to the electronic subsystem, must be driven by a tailored sequence of THz pulses. These pulses enable controlled constructive or destructive modulation of its amplitude. Second, this mode must couple nonlinearly and strongly to a Raman-active phonon, that eventually modifies the physical properties. Here, we realize this strategy in FePS$_3$, demonstrating the coherent control of a hidden metastable magnetic state via a double-THz-pulse excitation scheme \st{, and we identify the underlying nonlinear coupling pathway using two-dimensional THz spectroscopy}. While prior observations \cite{ilyas2024terahertz} established the existence of this THz-induced state, achieving coherent control over it was not an anticipated outcome. If the phase transition were driven by a conventional Raman process proportional to the square of the driving field, coherent control would be inefficient and limited to the brief duration of the THz pulses. Instead, using two-dimensional THz spectroscopy, we prove that the deterministic switching of the magnetization is governed by a two-phonon ionic pathway. Finding this specific mechanism provides the necessary physical framework to explain how macroscopic magnetic order can be controlled over extended timescales, moving beyond the observation of light-induced states to their active manipulation.

\subsection*{Principles of coherent control of magnetization in FePS$_3$}
\begin{figure} 
	\centering
	\includegraphics[width=\textwidth]{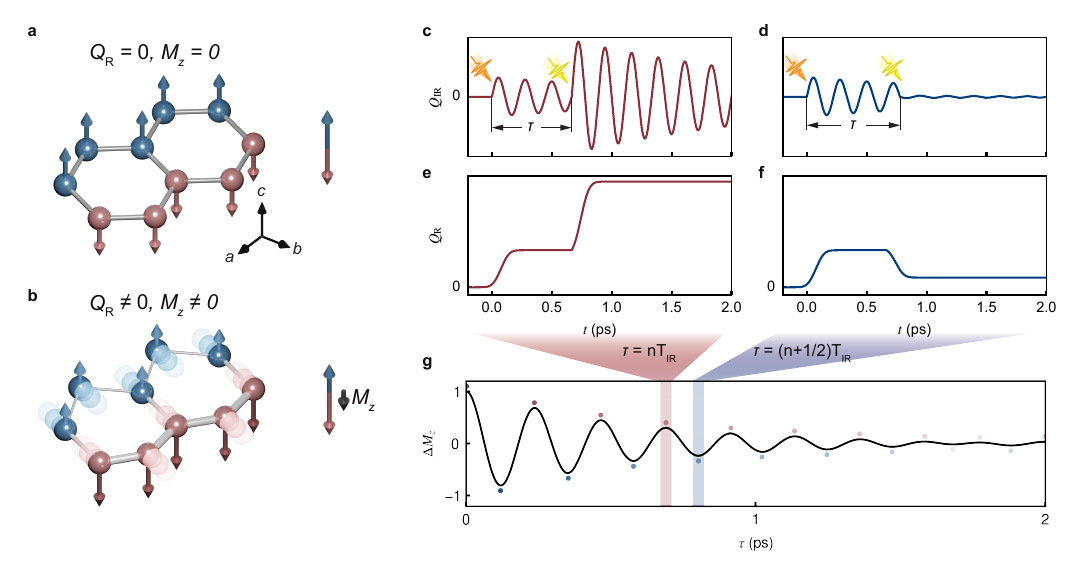} 
	\caption{\textbf{Principle of coherent phonon control of the metastable magnetization.}
		\textbf{a,} Equilibrium crystal and magnetic structure of FePS$_3$, with red and blue arrows indicating Fe spins pointing up and down in the out-of-plane direction, respectively. Right panel demonstrates fully compensated up and down spins. \textbf{b,} Crystal lattice displaced along the $\Omega_\text{R}= 3.27$~THz Raman phonon, showing enhanced exchange interactions within the red zigzag chain (thick bonds) and weakened exchange interactions within the blue zigzag chain (thin bonds). Right panel illustrates uncompensated up and down spins, generating net magnetization, $M_z$.
		\textbf{c-d,} Schematics of IR phonon amplitude as a function of time in double-pump experiment for $\tau=nT_\mathrm{IR}$ \textbf{c,} and $\tau=(n+\frac{1}{2})T_\mathrm{IR}$ \textbf{d}, demonstrating coherent enhancement \textbf{c} or suppression \textbf{d}. \textbf{e-f,} Corresponding Raman phonon displacement as a function of time resulting from IR phonon driving. \textbf{g,} Schematics of the THz-induced magnetization as a function of THz pulse separation $\tau$, with red and blue shadings correspond to $\tau=nT_\mathrm{IR}$ and $\tau=(n+\frac{1}{2})T_\mathrm{IR}$, respectively.}
	\label{fig:fig1} 
\end{figure}

FePS$_3$ exhibits exceptionally strong spin-lattice coupling among van der Waals (vdW) magnets \cite{liu2021direct, vaclavkova2021magnon, ergeccen2023coherent,luo2025terahertz,zong2023spin, zhou2022dynamical}, making it an ideal platform for exploring the control of magnetic order through structural tuning. The magnetic Fe$^{2+}$ ions are arranged in a hexagonal lattice, with spins forming anti-aligned zigzag ferromagnetic chains. Within this structure, spins point out of the plane due to strong magnetic anisotropy, resulting in Ising-like magnetic behavior~\cite{lee2016ising} (Fig. \ref{fig:fig1}a). Previous results have demonstrated that displacing the lattice along a specific Raman-active phonon mode at $\Omega_\text{R} = 3.27$~THz strengthens the nearest-neighbor magnetic bonds within one zigzag chain, while weakening interactions in the adjacent chain~\cite{ilyas2024terahertz}. Such modulation thereby alters the strength of magnetic exchange interactions in neighboring zigzag chains in an opposite way, leading to the emergence of a net magnetization out of the antiferromagnetic (AFM) motif (Fig. \ref{fig:fig1}b). Although both the macroscopic AFM ground state and the incident THz pulse preserve pure time-reversal symmetry ($\mathcal{T}$), a net magnetization emerges through a dynamic effect \cite{radaelli2018breaking}. Specifically, the THz-driven Raman structural distortion dynamically breaks the combined spatio-magnetic equilibrium symmetry ($\mathcal{T}' = \mathcal{T}\tau$, where $\tau$ is a lattice translation, see Supplementary Note 7), achieving a specific symmetry breaking that cannot be replicated by any macroscopic static strain or stress. Finally, while this initial structural drive decays within picoseconds, the induced magnetization persists for milliseconds because the lattice adiabatically follows the spin system, whose relaxation is heavily suppressed by critical slowing down near the Néel temperature \cite{ilyas2024terahertz} (Supplementary Note 8).


We can coherently control the metastable magnetization amplitude via three steps. First, by adjusting the timing between two nearly identical excitation pulses, the amplitude of an IR-active phonon can be significantly enhanced or suppressed. Specifically, when the time delay between pulses is an integer multiple of the phonon period, the forces from both pulses constructively interfere, amplifying the phonon amplitude (Fig. \ref{fig:fig1}c). Conversely, if the time delay corresponds to an odd integer multiple of half the phonon period, the forces destructively interfere, diminishing the phonon amplitude (Fig. \ref{fig:fig1}d). Second, if the IR-active phonon is anharmonically coupled to the Raman-active phonon at $\Omega_\text{R} = 3.27$~THz, this coupling induces a quasi-static distortion of the lattice along this Raman phonon eigenvector. Therefore, controlling the amplitude of the IR-active phonon will modify the displacement strength of the coupled Raman phonon (Fig. \ref{fig:fig1}e and Fig. \ref{fig:fig1}f). Third, as discussed earlier, the displacement of this specific phonon resonance in FePS$_3$ leads to the emergence of a new state with net magnetization. Through the process outlined above, the magnetization amplitude will be modulated at the frequency of the IR phonon as the time delay between pulses is varied (Fig. \ref{fig:fig1}g). Therefore, if experimentally realizable, coherent modulation of the IR-active phonons with resonant excitation would enable coherent control over the magnetization amplitude.  

\subsection*{Experimental results}
We employ time-dependent ellipticity change of a subsequent NIR pulse to probe the light-induced magnetization dynamics (Fig. \ref{fig:fig2}a) in transmission geometry. Near the magnetic ordering temperature ($T_\text{N}$ = 118~K), a single broadband THz pulse induces a new magnetic state with net magnetization, with its amplitude and lifetime dramatically increasing near the phase transition (Fig. \ref{fig:fig2}b). Notably, this magnetization can last for milliseconds close to $T_\text{N}$, thus forming a metastable state~\cite{ilyas2024terahertz}. To assess the feasibility of coherent phononic control of magnetization, we designed a set of different experiments. We first introduce a second THz pulse with a similar field profile (Fig. S1b) and continuously vary the time separation between the two pulses ($\tau$) near $\tau=0\ \mathrm{ps}$, while probing the induced dynamics at $t =~400$ ps when the system enters the metastable state (Fig. \ref{fig:fig2}c). The material’s response exhibits coherent oscillations as a function of $\tau$, with their amplitudes increasing as we approach $T_\text{N}$ (Fig. \ref{fig:fig2}d). A Fourier transform of these oscillations (Fig. \ref{fig:fig2}e) reveals a broad peak near ~4.5~THz, a frequency that is distinct from $\Omega_\text{R}$. Furthermore, whereas the metastable magnetization scales quadratically with the THz pulse field strength in the single-pump experiment (Fig. \ref{fig:fig2}f), indicating a nonlinear excitation mechanism, the oscillation amplitude scales linearly with the electric field strength of a single THz pulse (Fig. \ref{fig:fig2}g). These observations strongly suggest the involvement of an IR-active phonon that nonlinearly mediates the displacement of the Raman mode and thereby modulates the metastable magnetization. While a direct coupling between the driven infrared phonon and the macroscopic magnetization might intuitively seem possible, the Ginzburg-Landau analysis reveals that this direct pathway cannot produce the resonant critical enhancement near $T_\text{N}$. Consequently, the nonlinear rectification of the infrared drive into the Raman-active mode is the essential mechanism responsible for the long-lived magnetization observed experimentally. (see Supplementary Note 6).

We further examine the nature of the IR-active mode by studying the evolution of the oscillations as we simultaneously alter the polarization angle of the two THz pulses (Fig. \ref{fig:fig3}a). This polarimetry measurement allows for determination of the symmetry of the IR mode(s)~\cite{zhang2024terahertz,zhang2024terahertz-upconversion}, necessary for further illustrating their anharmonic interaction with $\Omega_\mathrm{R}$ the Raman-active mode. The Fourier spectra show a transfer of spectral weight between two frequency positions around 4.3 and 4.7 THz as we rotate the THz polarization (Fig. \ref{fig:fig3}b). The intensities of these two peaks maximize along orthogonal directions, implying the presence of two IR-active modes in close vicinity with perpendicular dipole moments. Indeed, the broad peak near 4.5~THz can be fit using two Lorentzians given in Eq. S7, whose weights change with polarization in an out-of-phase fashion (Fig. \ref{fig:fig3}c). These measurements indicate that the low- and high-frequency IR-active phonons carry electric dipole moments along the crystallographic $a$ and $b$ axes, and therefore follow $B_u$ and $A_u$ symmetries, respectively.

To investigate closely whether the IR phonon indeed mediates the displacement of the $\Omega_\mathrm{R}$ Raman mode, we performed two-dimensional THz spectroscopy (2DTS) measurements, a powerful technique for directly probing the nonlinear interactions between low-energy excitations and rectification processes \cite{mahmood2021observation, zhang2024terahertz,Johnson2019DistinguishingSpectroscopy} (see also Fig. S3). We conducted these measurements at $T=10$~K, where the larger amplitude and longer phonon coherence times compared to $T\approx T_\mathrm{N}$ allow for better examination of their interactions (see Fig. S4 for comparison). We continuously varied both the intra-pump time separation ($\tau$) and the pump-probe delay ($t$), whereas now $t$ is scanned over a short range ($-1$~ps $<t<10$~ps) with fine steps to resolve different modes (see Supplementary Text and Fig. S3). Such two-dimensional time scans ($\tau$ and $t$) are converted into two-dimensional frequency maps ($f_{\tau}$ and $f_t$) via Fourier transform (Fig. \ref{fig:fig3}d, Left). The displacement of the Raman mode is manifested in the 2D spectrum as the rectification signals near $f_t = 0$~THz. As shown in Fig.~\ref{fig:fig3}d, two strong spectral weight puddles appear near $f_{\tau}$ = 4.5~THz, in agreement with the frequencies extracted from the polarimetry measurements in Fig.~\ref{fig:fig3}c. These observations provide evidence for the involvement of the two nearly degenerate infrared phonons at ~4.5~THz in displacing the lattice and in dictating the coherent control of magnetization.

We repeated the 2DTS measurements at different polarization angles of both THz pulses to further elucidate the symmetry of the infrared modes displacing the lattice. The time traces and the corresponding frequency-frequency maps are provided in Fig. S3. The vertical linecuts at $f_{\tau}=0$~THz (Fig. \ref{fig:fig3}d, right) reveal the dominance of the mode near $f_t=4.3$~THz (4.7~THz) when the electric field is oriented along the crystallographic $a$-axis ($b$-axis), suggesting its $B_u$ ($A_u$) symmetry, in agreement with the results shown in Fig. \ref{fig:fig3}b and c. We note that the linewidths of these modes at $T=10$~K are much smaller than those near $T=118$~K, allowing for better resolution of the broad spectral peak.

\subsection*{Nonlinear phonon interactions}
The pathway connecting the IR-active phonons to the THz-induced magnetization is revealed by considering a theoretical model that includes all possible nonlinear couplings. We model the dynamics of the IR and Raman modes as coupled harmonic oscillators. The IR phonon $Q_\text{IR}$ is directly driven by the electric field of the THz pulse. In contrast, the Raman phonon $Q_\mathrm{R}$ can be driven by three different mechanisms up to the second order in electric field. First, the Raman phonon can be quadratically driven by the light field with the force proportional to $E_\text{THz}^2$~\cite{maehrlein2017terahertz}, where $E_\text{THz}$ represents the experimentally measured THz field. Second, known as ionic Raman scattering, it can be launched exclusively through nonlinear interactions with IR-active phonons, with the corresponding driving force proportional to $Q_\text{IR}^2$ \cite{forst2011nonlinear,juraschek2018sum}. The third possible mechanism, referred to as infrared resonant Raman scattering, involves both an IR phonon and the THz field \cite{khalsa2021ultrafast, mashkovich2021terahertz}, with the force proportional to $ E_\text{THz}Q_\text{IR}$. The latter two scenarios, involving the IR modes, are candidates for our coherent phononic control protocol.

The magnetization $M_z$ is modeled by an overdamped dynamics, driven by the Raman mode through the coupling term $gQ_\text{R}LM_z$~\cite{ilyas2024terahertz,vinasbostrom2025}, where $M_z$ is the magnetization magnitude, $L$ is the N\'eel vector magnitude, and $g$ is the coupling constant. We then derive the equations of motion for each variable (Eq. S1 and \cite{methods}) and numerically solve the dynamics of the IR phonon, the Raman phonon, and the magnetization by including two THz pulses reminiscent of the experiment setup (see Eq. S2). Next, we analyze the magnetization amplitude as we vary the time delay between THz pulses, $\tau$, while setting the probe pulse at a large delay time of $t=400$~ps. The simulation results in the time domain for each aforementioned scenario are provided in Fig. \ref{fig:fig4}a, with their Fourier transform shown in Fig. \ref{fig:fig4}b. Remarkably, in contrast to the other two scenarios, only the ionic Raman scattering channel exhibits a close resemblance to the experimental data, proving the dominance of the anharmonic phonon coupling in inducing the metastable magnetized state. 

While the above phenomenological model captures the essence of our experimental observations, we further performed atomistic spin dynamics simulations to account for critical fluctuations near $T_\mathrm{N}$. Using a $20\times 20$ supercell with the Fe$^{2+}$ spins initialized in thermal equilibrium through Monte Carlo simulated annealing (see \cite{methods} for details), we modeled the double-pump coherent control experiment with the ionic Raman scattering pathway activated. The resulting THz-induced magnetization $M_z(\tau)$ (Fig.~\ref{fig:fig4}c-d) shows a quantitative agreement with both the phenomenological model (Fig.~\ref{fig:fig4}a-b blue curve) and the experimental data (Fig.~\ref{fig:fig4}a-b black curve). Furthermore, first-principles phonon calculations in FePS$_3$ identify two IR-active phonon modes: a 4.34~THz mode with $B_u$ symmetry (Fig.~\ref{fig:fig4}e) and a 4.80~THz mode with $A_u$ symmetry (Fig.~\ref{fig:fig4}f). Their frequencies and corresponding symmetries closely align with our experimental assignments (Fig.~\ref{fig:fig3}c), further validating our interpretation.

\subsection*{Discussions and Outlook}
In summary, we have demonstrated coherent control of the nonequilibrium metastable magnetized phase in FePS$_3$ using resonant THz pulses. Our results extend the principles of coherent control in the field of femtochemistry \cite{baumert1991femtosecond, potter1992femtosecond} from molecular systems to crystalline materials. Notably, additional comparative double-pump experiments with near-infrared pulses (Fig. S5) showed no coherent modulation, underscoring the unique advantages of the THz approach. This remarkable selectivity of THz radiation over phononic and magnonic degrees of freedom promises access to a wide range of nonequilibrium phases of matter. While the specific $g L M Q_R$ coupling and zigzag magnetic order are unique to FePS$_3$, the underlying control mechanism is broadly applicable. Rectifying an infrared phonon into a Raman distortion alters local bond geometries. Because magnetic exchange interactions strongly depend on these structural parameters, this nonlinear phononic approach provides a general pathway to coherently control magnetic states in other systems with strong spin-lattice coupling. Practically, the ability to switch magnetization on and off with THz pulses effectively functions as an ultrafast quantum gate, enabling on-demand control with picosecond-scale response time that is orders of magnitude faster than conventional gigahertz quantum gates~\cite{kjaergaard2020superconducting,morgado2021quantum}. Looking ahead, leveraging materials with longer phonon coherence times, we may transform this platform into quantum memory systems, combining laser-based switching and magnetic readout. Our work thus establishes a fundamental framework for coherent terahertz manipulation of quantum materials, with promising implications for information processing technologies through precise control of emergent phases.\\

\newpage

\newpage

\begin{figure} 
	\centering
	\includegraphics[width=\textwidth]{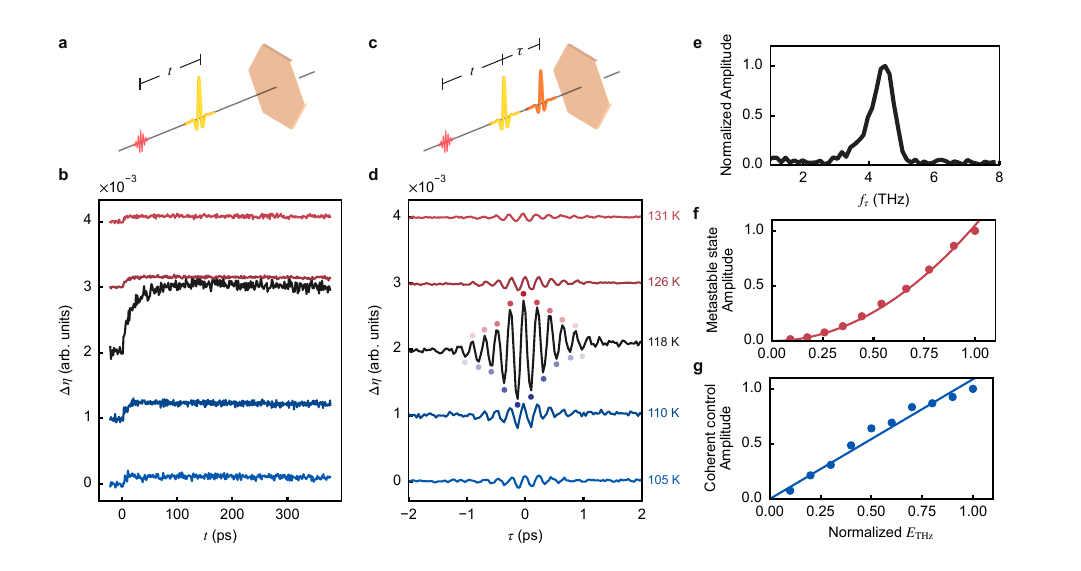} 
	\caption{\textbf{Experimental demonstration of coherent control.}
		 \textbf{a,} Schematic of the single THz pump experiment, with the yellow and red pulses representing THz pump and 800~nm probe. \textbf{b,} Ellipticity change time traces as a function of $t$ at different temperatures as measured in \textbf{a}. \textbf{c,} Schematics of the double THz pump experiment, with an additional THz pulse (orange) introduced at time delay $\tau$. \textbf{d,} Probe ellipticity change measured at $t=400\ \mathrm{ps}$ as a function of $\tau$ for various temperatures. \textbf{e,} Fourier transform of $\tau$-dependent trace at 118~K in \textbf{d}, revealing a peak at 4.5~THz. \textbf{f,} The metastable state magnitude in \textbf{b} as a function of normalized THz field strength $E_\mathrm{THz}$. The solid line is quadratic fitting. \textbf{g,} The area of the Fourier spectrum from \textbf{e} as a function of normalized THz field strength of the yellow THz pulse. The solid line is linear fitting.}
	\label{fig:fig2} 
\end{figure}

\begin{figure} 
	\centering
	\includegraphics[width=0.65\textwidth]{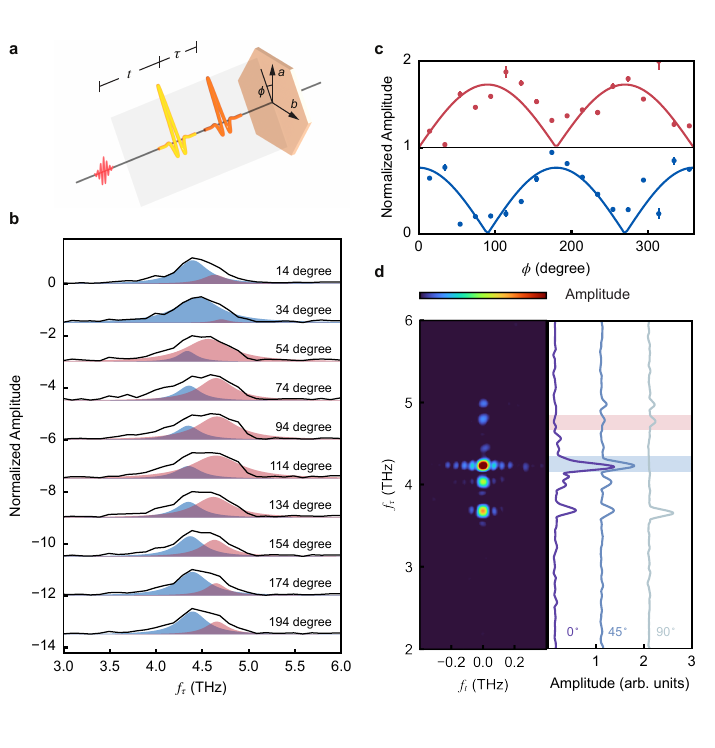} 
	\caption{\textbf{Identification of the IR-active phonons mediating the lattice displacement.}
		\textbf{a,} Schematic of the THz polarization dependence experiments, where the THz pulses are co-rotated. $\phi$ denotes the angle between THz polarization and the crystallographic $a$-axis. \textbf{b,} Evolution of the oscillation spectrum in Fig.~\ref{fig:fig2}e with $\phi$. The blue and red shaded areas are Lorentzian oscillator fits with Eq. S7. \textbf{c,} The amplitude of the two Lorentzian oscillators as a function of $\phi$, with $|\sin\phi|$ (red solid line) and $|\cos\phi|$ (blue solid line) fits demonstrating orthogonal polarizations. Data and fits are vertically offset for clarity. \textbf{d,} Left: 2D THz spectrum near $f_t=0\ \mathrm{THz}$ obtained at $T=10$~K and $\phi=45^\circ$. Right: Linecuts along $f_{\tau} =0$ THz of 2D THz spectra measured at $\phi=0^\circ,\ 45^\circ,\ 90^\circ$. The blue and red shaded regions correspond to the blue and red oscillator frequencies in \textbf{b} and \textbf{c}.}
	\label{fig:fig3} 
\end{figure}

\begin{figure} 
	\centering
	\includegraphics[width=\textwidth]{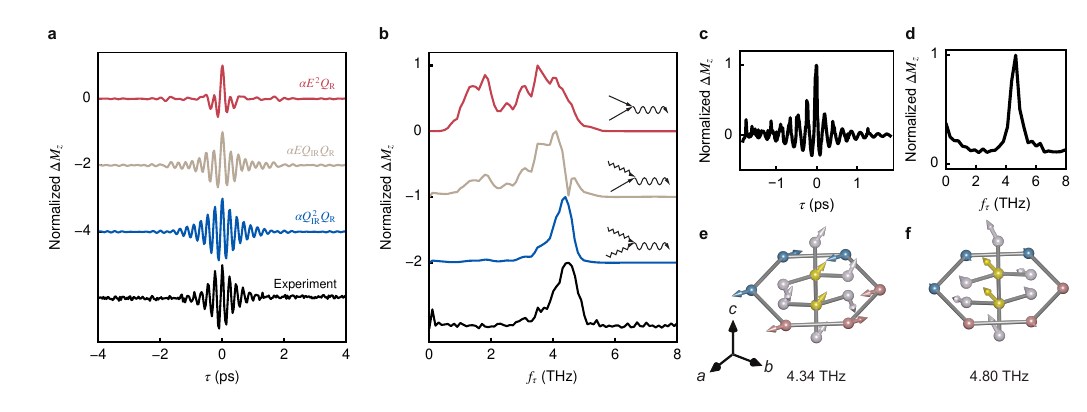} 
	\caption{\textbf{Possible nonlinear interaction pathways.}
	    \textbf{a,} Simulated $\Delta M(\tau)$ when two-photon excitation (red), infrared resonant Raman scattering (beige), and ionic Raman scattering (blue) mechanisms are activated for driving the Raman mode $\Omega_\text{R}$. The experimental result is shown in black. \textbf{b,} Fourier transform of the time traces in \textbf{a}. Inset shows the diagram of the nonlinear interactions, where the straight arrows represent photons and wavy arrows represent phonons. \textbf{c,} Atomistic spin dynamics simulation of $\Delta M(\tau)$ with the ionic Raman scattering mechanism activated. \textbf{d,} Fourier transform of \textbf{c}. \textbf{e-f,}. Schematics of the calculated IR phonon eigenmodes at 4.34~THz and 4.80~THz. The red and blue spheres represent Fe ions with spin pointing along opposite directions. The yellow and white spheres represent phosphorus and sulfur atoms, respectively.}
	\label{fig:fig4} 
\end{figure}
\clearpage


	


\clearpage 

%
\bibliography{NPstyle/coherent_control} 
\bibliographystyle{sciencemag}

%
%
%
%
%
%

\section*{Methods}
\subsubsection*{Sample preparation}
$\mathrm{FePS_3}$ single crystals were synthesized from iron (Sigma-Aldrich, 99.99\% purity), phosphorus (Sigma-Aldrich, 99.99\%), and sulfur (Sigma-Aldrich, 99.998\%) using the chemical vapor transport method. The powdered elements were prepared inside an argon-filled glove box. The starting materials were mixed in the stoichiometric ratio, with an additional 5 wt\% of sulfur to compensate for its high vapor pressure. We verified the stoichiometry of the synthesized single crystals with a COXEM-EM30 scanning electron microscope equipped with a Bruker QUANTAX 70 energy-dispersive X-ray system. The crystal structure was confirmed through X-ray diffraction measurements using a commercial diffractometer (Rigaku Miniflex II). Prior to experiments, single crystals were freshly cleaved along the [001] crystallographic direction to obtain clean surfaces.

\subsubsection*{Double-THz-pump experiments}
The experimental schematic of our double-THz-pump measurements is illustrated in Fig. S1a. The 800~nm output of a 1~kHz Ti:sapphire amplifier is split into two beams. The stronger beam pumps an optical parametric amplifier (OPA), whose signal and idler wavelength are set to 1450~nm and 1785~nm, respectively. These beams are modulated using optical choppers at half (500~Hz) and quarter (250 Hz) of the laser repetition rate to facilitate the detection of nonlinear signals through lock-in techniques~\cite{finneran2016coherent, Johnson2019DistinguishingSpectroscopy}. The temporal delay between the idler and the signal pulses is controlled by a mechanical delay stage (DS2). Both the signal and idler beams generate intense THz fields in a N-benzyl-2-methyl-4-nitroaniline (BNA) crystal through optical rectification process. The generated THz radiation is collected and focused onto the sample with a set of three gold-coated parabolic mirrors (PM1-3). The weaker 800 nm beam, delayed from the signal beam by DS1, is focused onto the sample overlapping with the THz beams. The THz-induced polarization ellipticity change of the transmitted probe beam is measured using the standard balanced detection technique. The time traces of the THz pulses generated by idler ($E_1$) and signal ($E_2$) are shown in Fig. S1b with the corresponding spectra shown in Fig. S1c. Both THz pulses exhibit broad spectral range spanning 0-6~THz.

\subsubsection*{First principles calculations}
To obtain the phonon modes FePS$_3$, we performed first principles simulations with the {\sc abinit} electronic structure code~\cite{Gonze2020TheDevelopments,Gonze1997First-principlesAlgorithm, Amadon2008Plane-waveOrbitals, Torrent2008ImplementationPressure}. We used the local density approximation with projector augmented wave (PAW) pseudopotentials, a plane wave cut-off of $20$ Ha and $40$ Ha respectively for the plane wave and PAW part, and included an empirical Hubbard $U$ of $2.7$ eV on the Fe $d$-orbitals as self-consistently determined in the {\sc octopus} electronic structure code via the ACBN0 functional. A $\Gamma$-centered Monkhorst-Pack grid with dimensions $8\times 6 \times 8$ was used to sample the Brillouin zone. The ground state was found to have zig-zag antiferromagnetic order with spins aligned along the $z$-axis.

Phonon frequencies and eigenvectors were calculated with {\sc abinit} after relaxing the atomic positions and stresses to below $10^{-6}$ Ha/Bohr. We find a relevant Raman phonon mode at $\Omega_{\rm R} = 3.27$ THz, as well as several IR active modes in the range $4 - 5$ THz. Out of the possible IR modes, we identify a $B_u$ mode at $4.34$ THz and a $A_u$ mode at $4.80$ THz, in good agreement with our experimental data.

\section*{Data availability}  
Datasets collected and/or analyzed during the current study are available from the corresponding author upon request.

\section*{Acknowledgments}
We thank Zhuquan Zhang and Keith Nelson for fruitful discussions and help with the experiments. 

\paragraph*{Funding:}
We acknowledge the support from the US Department of Energy, Materials Science and  Engineering  Division,  Office  of  Basic  Energy  Sciences  (BES  DMSE)  (data taking and analysis), Gordon and Betty Moore Foundation’s EPiQS Initiative grant GBMF9459 (instrumentation and manuscript writing). E.V.B. acknowledges funding from the European Union’s Horizon Europe research and innovation programme under the Marie Skłodowska-Curie Grant Agreement No. 101106809 (CavityMag). E.V.B. and A.R. acknowledge support from the Cluster of Excellence “Advanced Imaging of Matter” - EXC 2056 - project ID 390715994, SFB-925 “Light induced dynamics and control of correlated quantum systems” - project 170620586 of the Deutsche Forschungsgemeinschaft (DFG), from Grupos Consolidados (IT1453-22), and from the Max Planck-New York City Center for Non-Equilibrium Quantum Phenomena at the Flatiron Institute. The Flatiron Institute is a division of the Simons Foundation. The work at SNU was supported by the Leading Researcher Program of Korea’s National Research Foundation (Grant No. 2020R1A3B2079375).

\section*{Author contributions}
B.I., T.L. and N.G. conceived the study. B.I., T.L. and H.N. designed and built the experimental setup and performed the measurements. E.V.B. and A.R. performed the first-principles calculations and Monte Carlo simulations. J.P. and J.K. synthesized and characterized FePS$_3$ single crystals under the supervision of J.-G.P.; B.I., T.L. and H.N. performed the data analysis. B.I., T.L., H.N., A.v.H. and N.G. interpreted the data and wrote the paper with inputs from E.V.B., A.R., and all other authors. The project was supervised by N.G.

\section*{Competing interests}
The authors declare no competing interests.

\newpage


\renewcommand{\thefigure}{S\arabic{figure}}
\renewcommand{\thetable}{S\arabic{table}}
\renewcommand{\theequation}{S\arabic{equation}}
\renewcommand{\thepage}{S\arabic{page}}
\setcounter{figure}{0}
\setcounter{table}{0}
\setcounter{equation}{0}
\setcounter{page}{1} 


\begin{center}
\section*{Supplementary Materials for\\ \scititle}


	Batyr Ilyas$^{1,\dagger}$,
	Tianchuang Luo$^{1,\dagger}$,
    Honglie Ning$^{1,\dagger}$,
    Emil Viñas Boström$^{2, 3}$, \and
    Alexander von Hoegen$^{1}$, 
    Jaena Park$^{4}$, 
    Junghyun Kim$^{4}$,
    Je-Geun Park$^{4}$, 
    Angel Rubio$^{2, 3, 5}$,
	Nuh Gedik$^{1, \ast}$\and
\\ 
\small$^\ast$Corresponding author. Email: gedik@mit.edu\\
\small$^\dagger$These authors contributed equally to this work.
\end{center}

\subsubsection*{This PDF file includes:}
Supplementary Notes 1 to 7\\
Figures S1 to S5\\


\newpage


\section*{Supplementary Notes}





\section{Numerical simulation of coupled mode dynamics in double-THz-pump experiments}
To elucidate the dominant nonlinear coupling mechanism that drives the metastable state, we developed numerical simulations assuming that the IR phonon and Raman phonon are two driven oscillators and the light-induced magnetization exhibits an overdamped exponential decay. These three channels are coupled abided by symmetry and are described by three coupled differential equations:
\begin{equation}
\begin{gathered}
        \frac{d^2Q_\mathrm{IR}}{dt^2} + 2\Gamma_\mathrm{IR}\frac{dQ_\mathrm{IR}}{dt} + \omega_{\mathrm{IR}}^2Q_\mathrm{IR}  = Z^*E + 2\alpha_{11}Q_\mathrm{IR}Q_\mathrm{R}, \\
        \frac{d^2Q_\mathrm{R}}{dt^2} + 2\Gamma_\mathrm{R}\frac{dQ_\mathrm{R}}{dt} + \omega_{\mathrm{R}}^2Q_\mathrm{R}  = \alpha_{00} E^2 + \alpha_{01} EQ_\mathrm{IR} + \alpha_{11}Q^2_\mathrm{IR}, \\
        \frac{dM_z}{dt} = \Gamma_M(-2M_z+g Q_\mathrm{R}).
        \label{eq:nonlinearphonon}
\end{gathered}
\end{equation}
Here, $Q_\mathrm{IR}$ represents the IR-active mode with eigen-frequency $\omega_\mathrm{IR} = 2\pi (4.5\ \mathrm{THz})$ and damping coefficient $\Gamma_\mathrm{IR}=2\ \mathrm{THz}$. $Q_\mathrm{R}$ denotes the Raman-active mode with $\omega_\mathrm{R} = 2\pi (3.3\ \mathrm{THz})$ and damping coefficient $\Gamma_\mathrm{R}=20\ \mathrm{THz}$. These values are all acquired from the experiment. Only one of the IR-active modes is included in the simulation for simplicity. $Z^*$ is the effective charge, which reflects the oscillator strength of $Q_\mathrm{IR}$. $\alpha_{00}$, $\alpha_{01}$, and $\alpha_{11}$ are the excitation strengths of $Q_\mathrm{R}$ in the two-photon, one-photon-one-phonon, and two-phonon excitation channels. $M_z$ is the sample magnetization along the $c$-axis, which evolves with a very small phenomenological relaxation rate $\Gamma_M=10^{-4}\ \mathrm{THz}$. Finally, $g$ is the coefficient of the coarse-grained spin-phonon coupling term $gQ_\mathrm{R}LM_z$. As the value of the antiferromagnetic order parameter $L$ remains largely unchanged, its dynamics are ignored for simplicity.

\begin{figure} 
	\centering
	\includegraphics[width=\textwidth]{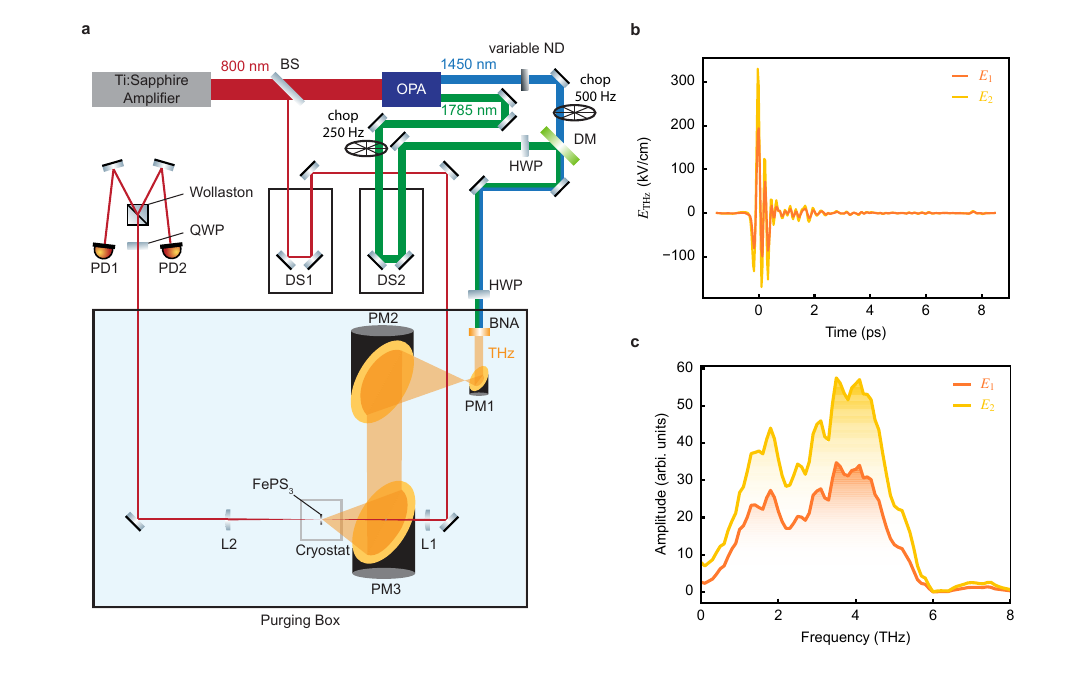}
	\caption{\textbf{Experimental configuration and THz pulse characterization.}
	\textbf{a,} Schematic diagram of the experimental setup. BS: beam splitter, ND: neutral-density filer, DM: dichroic mirror, HWP: half-wave plate, QWP: quarter-wave plate, L: lens, PD: photodiode, DS: delay stage, Wollaston: Wollaston prism. Unlabeled optics represent mirrors.
        \textbf{b,}	THz electric field as a function of time for $E_1$ (generated by 1785 nm pump) and $E_2$ (generated by 1450 nm pump).
        \textbf{c,} Fourier transform of the time traces in \textbf{b}.
        }
	\label{fig:sup_setup}
\end{figure}

The driving field $E$ is adopted from the experimentally determined THz field time traces $E_1$ and $E_2$ as depicted in Fig.~\ref{fig:sup_setup}b: 
\begin{equation}
    E(t) = E_1(t+\tau)+E_2(t).
    \label{eq:THzfield}
\end{equation}
The temporal evolution of $Q_\mathrm{IR}$, $Q_\mathrm{R}$, and $M_z$ over $t$ is computed numerically with a 4th-order Runge-Kutta method over the time range $-3\ \mathrm{ps}<t<200\ \mathrm{ps}$ and intra-pump delay $-5\ \mathrm{ps}<\tau<5\ \mathrm{ps}$ to emulate the experiment. Additional simulations are run with $E_1$ and $E_2$ set to zero, respectively, to obtain the nonlinear coherent modulation component~\cite{lu2017coherent}
\begin{equation}
\Delta M_z(t,\tau) = M_z^{E_1,E_2}(t,\tau)-M_z^{E_1=0,E_2}(t,\tau)-M_z^{E_1,E_2=0}(t,\tau).
\end{equation}
Finally, we plot the saturation value of $\Delta M_z$ at large $t$ ($\Delta M_z(t=400\ \mathrm{ps})$) as a function of $\tau$, which is shown in Fig.~\ref{fig:fig2}d.

\section{Monte Carlo simulations and atomistic spin dynamics}
To supplement the Ginzburg-Landau dynamics, we performed real time simulations of the atomistic spin system. For this purpose we consider the spin-phonon Hamiltonian
\begin{align}
 H &= \sum_{ij} {\bf S}_i \cdot ({\bf J}_{ij} {\bf S}_j) - \Delta \sum_i (S_i^z)^2 + \frac{1}{2} \sum_{i} \Big[ P_i^2 + \Omega_i^2 Q_i^2 \Big] + \gamma Q_{\rm R} \sum_{ij} A_{ij} {\bf S}_i \cdot {\bf S}_j \nonumber \\
 &+ gQ_{\rm R}  Q_{\rm IR}^2 + Z E(t) Q_{\rm IR},
\end{align}
including two phonon modes at $3.27$ THz and $4.5$ THz (the Raman active mode and one of the IR-active modes) and the external THz electric field $E(t)$. The $4.5$ THz mode is considered to be linearly driven, with a light-matter coupling given by the Born effective charge $Z$. The Hamiltonian leads to the following coupled equations of motion
\begin{align}
 \frac{\partial {\bf S}_i}{\partial t} &= {\bf S}_i \times \frac{\partial H}{\partial {\bf S}_i} \nonumber \\
 \frac{\partial^2 Q_{\rm R}}{\partial t^2} &= -\Omega_{\rm R}^2 Q_{\rm R} - g Q_{\rm IR}^2 - \gamma \sum_{ij} A_{ij} {\bf S}_i \cdot {\bf S}_j \\
 \frac{\partial^2 Q_{\rm IR}}{\partial t^2} &= - \Omega_{\rm IR}^2 Q_{\rm IR} - 2g Q_{\rm R}  Q_{\rm IR} + Z E(t). \nonumber
\end{align}
For the electric field we use the experimental pulses, separated by a delay $\tau$.

The equilibrium state of the magnetic system at a temperature $T = T_{\rm N}$ was prepared with simulated annealing using the Monte Carlo Metropolis algorithm. All exchange parameters have been calculated from first principles, and are given in Ref.~\cite{ilyas2024terahertz}. The calculations were performed for a monolayer sample of FePS$_3$ using a spin length $S = 2$ and an in-plane supercell of linear size $L = 20$ unit cells. The simulated annealing was initialized at a temperature $T \approx 170$ K and performed down to a target temperature $T = T_{\rm N} = 115$ K in steps of $\Delta T \approx 2$ K. At each temperature we performed $2000$ Monte Carlo sweeps to thermalize the system, followed by $4000$ measurements performed at an interval of $50$ sweeps to obtained thermodynamic averages. Our calculations find a magnetic transition temperature in good agreement with experiment, thereby validating the calculated equilibrium spin parameters.

At the transition temperature, the statistical average of the magnetization, $M_z = \langle \sum_i S_i^z\rangle$, is found to vanish. To investigate its dynamics, we then apply the THz field $E(t) = E_1(t+\tau) + E_2(t)$ assuming a Born effective charge $Z = 8$ (from $4$ Fe ions each with an effective charge $Z_0 = 2$) and a phonon-phonon coupling $g = 1$ meV. The time evolution was performed up to a maximal time $t = 662$ ps, using $N_t = 20000$ time steps and a step size of $dt = 0.033$ ps. The magnetization was evaluated as a function of $\tau$, at a time $t_{\rm eval} = 200$ ps after the pulse $E_2$, when the magnetization has reached a saturated value. The resulting magnetization is shown in Fig.~\ref{fig:fig4}c, and its Fourier transform in Fig.~\ref{fig:fig4}d.



\section{Polarization dependence of the coherent control of metastable states}

\begin{figure}
	\centering
	\includegraphics[width=0.65\textwidth]{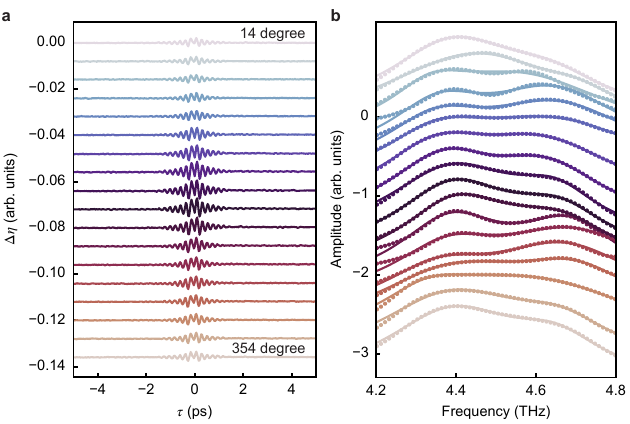}
	\caption{\textbf{THz polarization dependence of the coherent modulation.}
	\textbf{a,} Coherent modulation amplitude as a function of two-THz delay $\tau$ at various THz polarizations. \textbf{b,} Fourier transform of \textbf{a} and fitting with Eq.~\ref{eq:lorentz}.
        }
	\label{fig:sup_pol_time}
\end{figure}

The modulation of the metastable state as a function of time delay $\tau$ between the double THz pulses for various THz polarizations is shown in Fig.~\ref{fig:sup_pol_time}. The THz polarization of both $E_1$ and $E_2$ are simultaneously rotated from $\phi=14^\circ$ to $\phi=374^\circ$ with $20^\circ$ increments. Their corresponding spectra as shown in Fig.~\ref{fig:sup_pol_time}b and Fig.~\ref{fig:fig3}b is obtained through Fourier transforming the time traces in Fig.~\ref{fig:sup_pol_time}a. The transformed spectra are then fitted with two complex Lorentzian functions \cite{luo2025terahertz}, because a coherent oscillation can be modeled as a damped harmonic oscillator with maximum amplitude at $\tau=0$:
\begin{equation}
    f(\tau) = A \cos(2\pi f_0\tau+\varphi)e^{-\gamma|\tau|},
    \label{eq:oscillation}
\end{equation}
where $A$ represents the maximum amplitude, $f_0$ is the oscillation frequency, $\phi$ denotes the oscillation phase, and $\gamma$ is the coherence damping rate. The corresponding Fourier transform of Eq.~\ref{eq:oscillation} is given by the addition of two Lorentzians
\begin{equation}
    \tilde{f}(\omega) = A\gamma \left[\frac{e^{i\varphi}}{\gamma^2+(\omega-2\pi f_0)^2}+\frac{e^{-i\varphi}}{\gamma^2+(\omega+2\pi f_0)^2}\right].
    \label{eq:lorentz}
\end{equation}
Since our analysis focuses exclusively on the positive frequency domain, we can neglect the second term in Eq.~\ref{eq:lorentz} and therefore the Fourier spectrum of each IR phonon can be modeled by a Lorentzian. The polarization-dependent amplitude of the two modes (Fig.~\ref{fig:fig3}c) is extracted by fitting both the real and imaginary parts of the spectrum with a sum of two oscillators described by Eq.~\ref{eq:lorentz}. The fitted results are shown in Fig.~\ref{fig:sup_pol_time}b.

\section{2D THz spectroscopy of the coherent modes in FePS$_3$}
\begin{figure}
	\centering
	\includegraphics[width=\textwidth]{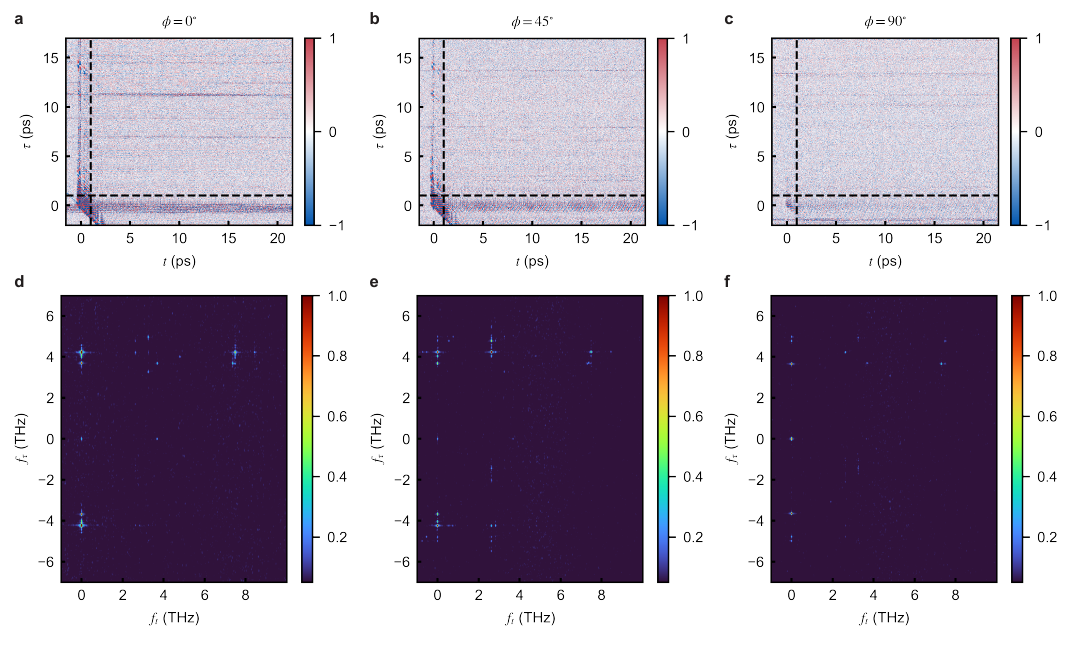}
	\caption{\textbf{2D THz signal as a function of time delays $(t,\tau)$.}
	\textbf{a-c,} 2D THz response measured at THz polarization angles $\phi=0^\circ$, $45^\circ$, and $90^\circ$. The black dashed lines demarcate the region used for two-dimensional Fourier transformation analysis. The signal strength is represented in arbitrary units.
    \textbf{d-f,} Fourier transform of the selected regions in \textbf{a-c}.
        }
	\label{fig:2dthz_time}
\end{figure}

The 2D THz responses of FePS$_3$ as a function of $(t,\tau)$ measured at 10~K are presented in Fig.~\ref{fig:2dthz_time}. The measurements are performed at various THz polarization angles $\phi$. To obtain the frequency-domain spectra shown in Fig.~\ref{fig:fig3}d, we perform a 2D Fourier transform over the region bounded by the black dashed lines in the upper-right quadrant of the time-domain data. This region is selected to maximize the signal-to-noise ratio while avoiding temporal artifacts from the pump-pump and pump-probe overlap region.

The complete 2D THz spectra are presented in Fig.~\ref{fig:2dthz_time}d-f. The horizontal axis ($f_t$) represents the detection frequency, while the vertical axis ($f_\tau$) corresponds to the excitation frequency. 
The sharp peaks observed in these spectra can be classified into the following categories~\cite{zhang2024terahertz}:
\begin{enumerate}
    \item \textbf{Nonlinear coupling peaks ($f_\tau\neq f_t \neq 0$):} These off-diagonal peaks indicate nonlinear coupling between IR-active phonons at frequencies $f_\tau$ and Raman-active phonons at frequencies $f_t$. They can originate from either two-phonon or one-phonon-one-photon processes.
    \item \textbf{THz rectification peaks ($f_t=0$):} They represent displacements generated from two-phonon or two-magnon nonlinearities.
    \item \textbf{Magnon nonlinear peaks} ($f_t = f_M$ or $f_\tau = f_M$. $f_M=3.7\ \mathrm{THz}$ is the magnon frequency)\textbf{:} These peaks, which include $(f_t,f_\tau)=(f_M,f_M), (f_M,0),(0,f_M),(2f_M,f_M)$ arise due to magnon anharmonicities.
\end{enumerate}

Notably, THz rectification peaks at $f_t=4.2\ \mathrm{THz}$ and $4.8\ \mathrm{THz}$ are observed (Fig.~\ref{fig:fig3}d), indicating their role in creating lattice displacement as discussed in the main text. 

\section{Absence of coherent modulation with double NIR pumps in FePS$_3$}
To demonstrate that THz pulses are crucial for coherent control of the metastable state, we carried out double-pump experiments with photon energies in the near-infrared (NIR) range in FePS$_3$ at 115~K. First, we pump FePS$_3$ with a single NIR pulse at 1450~nm. Ultrafast demagnetization is detected in the polarization rotation channel $\Delta\theta$ (Fig.~\ref{fig:2dnir}a). We then added a second NIR pulse at 1785~nm and scanned the time delay between the two NIR pump $\tau$ with $t$ fixed to $370\ \mathrm{ps}$. Only a sharp enhancement of the signal is observed when the two pumps overlap, indicating the absence of coherent modulation with NIR pumps (Fig.~\ref{fig:2dnir}b).

\section{Analysis of direct infrared phonon-induced magnetization}
In this section, we consider the possibility of a direct coupling between the infrared (IR) phonons and the magnetic order based on effective Ginzburg-Landau theory.

First, the equilibrium ground state preserves the combined spatio-magnetic symmetry $\mathcal{T}'=\mathcal{T}\tau$, where $\mathcal{T}$ is the time-reversal operator, and $\tau$ is lattice translation. Under this operation, the product $LM$ is odd ($L\xrightarrow{\text{$\mathcal{T'}$}} L$, $M\xrightarrow{\text{$\mathcal{T'}$}} -M$ ), where $L$ is the zig-zag antiferromagnetic order parameter, $M$ is the ferromagnetic magnetization. Thus, to form an invariant free energy term, $LM$ must couple to a linear structural coordinate that is also odd under $\mathcal{T}'$ (such as the zone-folded Raman mode $Q_\text{R}$). In contrast, a directly driven IR phonon ($Q_\text{IR}$) must enter the free energy quadratically ($Q_\text{IR}^2$) to respect inversion symmetry, since IR modes are odd under spatial inversion while magnetic moments ($L$, $M$) are even pseudo-vectors. However, any squared real coordinate is strictly even under $\mathcal{T'}$. Therefore, the proposed coupling term $Q_{IR}^2LM$ remains odd under $\mathcal{T'}$ and is forbidden by the equilibrium magnetic space group (i.e., the coupling constant $g_\text{IR}=0$).

We evaluate the direct coupling between infrared-active phonons and macroscopic magnetization within the framework of an effective Ginzburg-Landau theory, and using an effective quadratic coupling $g_\text{IR}Q_\text{IR}^2LM$. We assume the free energy is of the form~\cite{ilyas2024terahertz,vinasbostrom2025}
\begin{align}
 F &= \frac{a_L(T)}{2} L^2 + \frac{b_L}{4} L^4 + \frac{a_M(T)}{2} M^2 + \frac{b_M}{4} M^4 \\
 &+ \frac{\Omega_{\rm R}}{2} Q_{\rm R}^2 + \frac{\Omega_{\rm IR}}{2} Q_{\rm IR}^2 + g_{\rm R} (L - \langle L\rangle) MQ_{\rm R} + g_{\rm IR} L M Q_{\rm IR}^2 + F_{\rm R} Q_{\rm IR} + F_{\rm IR} Q_{\rm IR}. \nonumber
\end{align}
Here $\Omega_{\rm R}$ and $\Omega_{\rm IR}$ are frequencies of Raman (R) and infrared (IR) phonons. We have added the force terms $F_{\rm R}$ and $F_{\rm IR}$ coupling to the Raman and infrared phonons, for reasons that will become clear below.

To find the minima of the free energy we assume $L \gg M$ and $Q$, and differentiate $F$ with respect to the later variables. This gives
\begin{alignat}{2}
 0 &= \frac{\partial F}{\partial M} &&= a_M M + b_M M^3 + g_{\rm R} (L - \langle L\rangle)Q_{\rm R} + g_{\rm IR} LQ_{\rm IR}^2 \\
 0 &= \frac{\partial F}{\partial Q_{\rm R}}  &&= \Omega_{\rm R}  Q_{\rm R}  + g_{\rm R} (L - \langle L\rangle)M + F_{\rm R} \nonumber \\
 0 &= \frac{\partial F}{\partial Q_{\rm IR}} &&= \Omega_{\rm IR} Q_{\rm IR} + 2g_{\rm IR} LM Q_{\rm IR} + F_{\rm IR}. \nonumber
\end{alignat}
We note that due to the quadratic coupling to $Q_{\rm IR}$, the last equation implies that $Q_{\rm IR} = 0$ for $F_{\rm IR} = 0$ (neglecting the fine tuning condition $\Omega_{\rm IR} + 2g_{\rm IR} LM = 0$). With a finite force term, we can solve both equations for the phonon coordinates to obtain
\begin{align}
 Q_{\rm R} &= - \frac{g_{\rm R} (L - \langle L\rangle)M}{\Omega_{\rm R}} - \frac{F_{\rm R}}{\Omega_{\rm R}}\\
 Q_{\rm IR} &= -\frac{F_{\rm IR}}{\Omega_{\rm IR} + 2g_{\rm IR} LM}. \nonumber
\end{align}
We see that these couplings are of a very different form, such that phonon displacements are enhanced (suppressed) for large $L$ for the Raman (infrared) phonon.

We now use these solutions in the equation for the magnetization, and find
\begin{align}
 a_M M + b_M M^3 &= \frac{g_{\rm R}^2 (L - \langle L\rangle)^2}{\Omega_{\rm R}} M + \frac{g_{\rm R} F_{\rm R} (L - \langle L\rangle)}{\Omega_{\rm R}} - \frac{g_{\rm IR} F_{\rm IR}^2 L}{(\Omega_{\rm IR} + 2g_{\rm IR} LM)^2}.
\end{align}
Considering these terms separately, we find the coupling to the Raman mode (to lowest order in $F_{\rm R}/\Omega_{\rm R}$) gives a magnetization of the form
\begin{align}
 M_{\rm R} &= \Big( \frac{g_{\rm R} F_{\rm R}}{\Omega_{\rm R} L_s} \Big) \frac{\chi_L}{Va_M - (g_{\rm R}^2/\Omega_{\rm R}) \chi_L} \approx \Big( \frac{g_{\rm R} F_{\rm R}}{\Omega_{\rm R} a_M} \Big) \frac{\chi_L}{V L_s},
\end{align}
where $L_s$ is the zig-zag saturation magnetization, and we have assumed $F_R = F_R L/L_s$ to account for the fact that $Q_R$ is defined relative to the sign of $L$~\cite{ilyas2024terahertz,vinasbostrom2025}. The dependence on the susceptibility $\chi_L = V (\langle L^2\rangle - \langle L\rangle^2) \sim |T - T_\textrm{N}|^{-\gamma}$ shows the resonant response of $M$ to any external for $F_R$ for $T \approx T_{\rm N}$. 

For the infrared phonon, we similarly get
\begin{align}
 b_M M_{\rm IR}^3 &= - \frac{g_{\rm IR} F_{\rm IR}^2 L}{(\Omega_{\rm IR} + 2g_{\rm IR} LM_{\rm IR})^2} - a_M M_{\rm IR}
\end{align}
Assuming $M_{\rm IR} \ll 1$, we can linearize this equation to find
\begin{align}
 M_{\rm IR} &= \frac{g_{\rm IR} F_{\rm IR}^2}{\Omega_{\rm IR}^2} \frac{L}{(2g_{\rm IR}^2 F_{\rm IR}^2/\Omega_{\rm IR}^4) L^2 - a_M} \approx - \frac{g_{\rm IR} F_{\rm IR}^2}{\Omega_{\rm IR}^2} \frac{L}{a_M}.
\end{align}
Although this coupling also gives a finite magnetization for $T < T_{\rm N}$, there is no resonant enhancement close to the N\'eel temperature, and the effect is a factor $F_{\rm IR}/\Omega_{\rm IR}$ smaller than the Raman coupling. 

Therefore, our thermodynamic modeling reveals that a direct IR pathway cannot produce the observed metastable state, making the nonlinear rectification into the Raman mode physically necessary. 

We also note that the specific symmetry of the Raman phonon-induced distortion cannot be achieved by global static strain or stress. Static strain modifies the Bravais lattice vectors through an affine deformation, leading to a uniform stretching, compression, or shear of all symmetry-related bonds. By contrast, the relevant Raman mode identified in our study involves a specific internal coordinate shift where the bonds along one zig-zag chain are strengthened, while those in the adjacent chain are weakened. This staggered deformation breaks the equivalence of the antiferromagnetic sublattices and generates a net magnetization. Such alternating bond modulation cannot be generated by global uniaxial, biaxial, or hydrostatic strain, which act uniformly on the lattice.

\section{Symmetry analysis of THz-induced magnetization}
At first glance, the emergence of magnetization may seem counterintuitive, as both the macroscopic magnetic state of FePS$_3$ and the linearly polarized THz pulse preserve time-reversal symmetry. To clarify the magnetic symmetry of the equilibrium ground state: while the local AFM order breaks pure time-reversal symmetry ($\mathcal{T}$), the macroscopic AFM state preserves time reversal symmetry. More precisely, the ground state is invariant under the combined operation of time reversal and a lattice translation, which we denote as $\mathcal{T'}=\mathcal{T}\tau$. It is this combined symmetry that forbids a net macroscopic magnetization ($M=0$) in equilibrium.

The emergence of a macroscopic magnetization under linearly polarized terahertz drive is a consequence of a dynamical coupling, governed by the trilinear free energy term $\mathcal{F}\propto gQ_\text{R}LM$, where $L$ is the AFM order parameter, $M$ is the net magnetization, and $Q_\text{R}$ is the Raman phonon displacement.

This coupling term is allowed by the equilibrium magnetic space group of FePS$_3$. Under the symmetry operation $\mathcal{T'}$:
\begin{itemize}
    \item The AFM order parameter $L$ is invariant by definition ($L\xrightarrow{\text{$\mathcal{T'}$}} L$ ).
    \item The macroscopic magnetization $M$ is odd under time reversal but invariant under spatial translation ($M\xrightarrow{\text{$\mathcal{T'}$}} -M$).
    \item The zone-folded Raman phonon QR, which involves an out-of-phase motion between adjacent structural layers, is invariant under time reversal but odd under the lattice translation  ($Q_\text{R}\xrightarrow{\text{$\mathcal{T'}$}} -Q_\text{R}$ ).
\end{itemize}

Consequently, the product transforms as $Q_\text{R}LM\xrightarrow{\text{$\mathcal{T'}$}} Q_\text{R}LM$, respecting the equilibrium symmetry.

The mechanism of symmetry breaking in the driven system is as follows. The linearly polarized THz pulse displaces the lattice, and induces a finite $Q_\text{R}$. This lattice displacement dynamically breaks the combined $\mathcal{T'}$ symmetry of the ground state. The displaced phonon acts as an effective, internal magnetic field $H_\text{eff}\propto gQ_\text{R}L$, which explicitly breaks the $\mathcal{T'}$ symmetry and induces a finite macroscopic magnetization $M$. In this sense, the linearly polarized light does not break time-reversal symmetry directly, rather, it drives a structural distortion that dynamically breaks the spatio-magnetic symmetry ($\mathcal{T'}$) protecting the $M=0$ state.

\section{Spin-lattice relaxation dynamics near $T_N$}
To explain the difference between the tens of picosecond duration of the lattice modes and the millisecond lifetime of the induced magnetization, in this section we outline the relevant thermodynamic and critical relaxation mechanisms. More extended analysis and simulations are provided in Ref. \cite{ilyas2024terahertz}.

The metastable state is initiated by a THz-driven coherent phonon displacement. Time-resolved measurements show that these initial large-amplitude phonon oscillations indeed decay within tens of picoseconds \cite{ilyas2024terahertz, luo2024terahertz}. However, because the structural and magnetic degrees of freedom are coupled via the trilinear free energy term $\mathcal{F}\propto gQ_\text{R}LM$, the metastable magnetization $M$ exerts a continuous force on the lattice, given by $F_\text{eff}=-\frac{\partial\mathcal{F}}{\partial Q_\text{R}}\propto gLM$.  Consequently, the spin system maintains a small, quasi-static structural displacement for its entire duration. The lattice adiabatically follows the spin system and the millisecond lifetime is therefore dictated by the spin relaxation.

The relaxation of the spin system back to the $M=0$ ground state is governed by critical dynamics near the Néel temperature $T_\text{N} \approx120$ K. In this regime, the magnetic correlation length  diverges. Following the theory of dynamic critical phenomena \cite{Hohenberg1977}, the characteristic relaxation time $\tau$ scales as $\tau\propto \chi \sim |T-T_\text{N}|^{\nu z}$ (where $\nu$ and $z$ are the dynamic critical exponents).  This macroscopic critical slowing down suppresses the spin relaxation rate, extending the lifetime of the metastable magnetization into the millisecond regime, which is orders of magnitude longer than typical picosecond spin dynamics.

Finally, we note that a full ab-initio description of the phenomena discussed in this work is limited because the mechanism fundamentally depends on thermal fluctuations. Specifically, the photo-induced magnetization is only observed close to the Néel temperature, and vanishes at both higher and lower temperatures. Because first-principles methods like time-dependent density functional theory (TD-DFT) are restricted to zero temperature, they cannot capture these essential effects. In addition, simulating the long timescales needed to describe the phonon dynamics is computationally prohibitive for these methods. To explain these dynamics, we instead rely on a first-principles-based effective model \cite{ilyas2024terahertz}, which shows that the Raman phonon at 3.27 THz mediates a coupling between the magnetization $M$ and the fluctuations of the antiferromagnetic order $L$.

\newpage


\newpage

\begin{figure}
	\centering
	\includegraphics[width=0.5\textwidth]{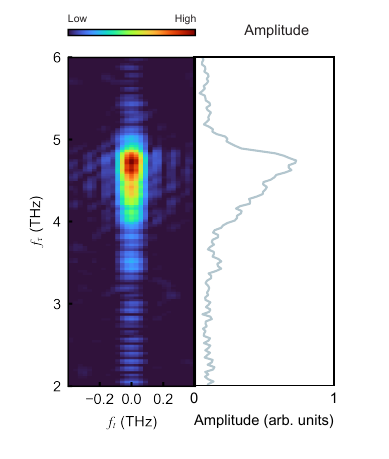}
	\caption{\textbf{2D THz spectrum around $f_t=0\ \mathrm{THz}$ obtained near $T_\mathrm{N}$.}
	Left: 2D THz spectrum near $f_t=0\ \mathrm{THz}$ obtained at $T=118$~K. Right: Linecut along $f_{\tau} =0$ THz of the 2D THz spectra. The peak corresponding to $Q_\mathrm{IR}$ exhibits thermal broadening compared to the low temperature results (the blue and red shaded regions in Fig.~\ref{fig:fig3}d).
        }
	\label{fig:2dthz_118}
\end{figure}

\begin{figure}
	\centering
	\includegraphics[width=0.65\textwidth]{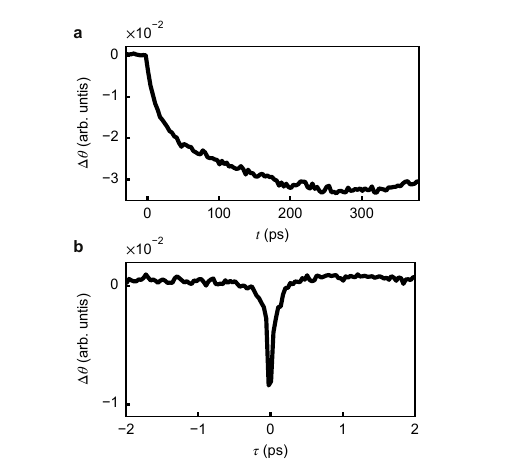}
	\caption{\textbf{Double NIR pump demagnetization response in FePS$_3$ at 115~K.}
	\textbf{a,} Single NIR pump induced demagnetization as a function of pump-probe delay $t$. \textbf{b,} Double NIR pump modulation of demagnetization as a function of $\tau$ measured at $t=370\ \mathrm{ps}$.
        }
	\label{fig:2dnir}
\end{figure}

\clearpage
\end{document}